\documentclass[11pt]{iopart}
\pdfoutput=1
\usepackage{cite,amssymb,amsfonts,graphicx,bm,dcolumn,mathrsfs}
\usepackage{color}
\usepackage{ulem}
\usepackage[utf8]{inputenc}
\usepackage[T1]{fontenc}
\usepackage{lmodern}

\begin{document}

\newcommand{\dif}{\rmd}
\newcommand{\ee}{\rme}
\newcommand{\ii}{\rmi}
\newcommand{\kB}{k_\mathrm{B}}
\newcommand{\sub}[1]{\mathrm{#1}}
\newcommand{\vect}[1]{\bi{#1}}

\title{Modified Thirring model beyond the excluded-volume approximation}
\vspace{5mm}
\author{Alessandro Campa$^1$, Lapo Casetti$^{2,3}$, Pierfrancesco Di Cintio$^{2,3,4}$, Ivan Latella$^5$, J. Miguel Rubi$^5$
and Stefano Ruffo$^{4,6}$}
\address{$^1$ National Center for Radiation Protection and Computational Physics, Istituto Superiore di Sanit\`{a},
Viale Regina Elena 299, 00161 Roma, Italy}
\address{$^2$ Dipartimento di Fisica e Astronomia, Universit\`a di Firenze,\\and INFN, Sezione di Firenze,
via G.\ Sansone 1, 50019 Sesto Fiorentino (FI), Italy}
\address{$^3$ INAF-Osservatorio Astrofisico di Arcetri, Largo E.\ Fermi 5, 50125 Firenze, Italy}
\address{$^4$ Istituto dei Sistemi Complessi, Consiglio Nazionale delle Ricerche, Via Madonna del Piano 10, 50019 Sesto Fiorentino, Italy}
\address{$^5$ Departament de F\'isica de la Mat\`eria Condensada, Universitat de Barcelona, \\ Mart\'i i Franqu\`es~1,
08028 Barcelona, Spain}
\address{$^6$ SISSA, via Bonomea 265 and INFN, Sezione di Trieste, 34136 Trieste, Italy}
\ead{\mailto{alessandro.campa@iss.it}, \mailto{lapo.casetti@unifi.it}, \mailto{pierfrancesco.dicintio@cnr.it}, 
\mailto{ilatella@ub.edu}, \mailto{mrubi@ub.edu} and \mailto{ruffo@sissa.it}}

\begin{abstract}
Long-range interacting systems may exhibit ensemble inequivalence and can possibly attain equilibrium states under completely open
conditions, for which energy, volume and number of particles simultaneously fluctuate. Here 
we consider a modified version of the Thirring model for self-gravitating systems with attractive and repulsive long-range
interactions in which particles are treated as hard spheres in dimension $d=1,2,3$. Equilibrium states of the model are studied
under completely open conditions, in the unconstrained ensemble, by means of both Monte Carlo simulations and analytical methods
and are compared with the corresponding states at fixed number of particles, in the isothermal-isobaric ensemble. Our theoretical
description is performed for an arbitrary local equation of state, which allows us to examine the system beyond the excluded-volume
approximation. The simulations confirm the theoretical prediction of the possible occurrence of first-order phase transitions in
the unconstrained ensemble. This work contributes to the understanding of long-range interacting systems exchanging heat, work and
matter with the environment.
\end{abstract}


\maketitle

\section{Introduction}

Systems with long-range interactions are characterized by a slowly-decaying interaction potential that couples particles at relatively
large distances, comparable to the system size. In particular, interactions decaying with a power smaller than the space dimension
$d$ are long-range even if the system size goes to infinity. The study of these systems has attracted considerable attention in recent
years because of several properties that are absent when the interactions are short-range, both under equilibrium and non-equilibrium
conditions~\cite{Campa_2014,Campa_2009,Levin_2014,Bouchet_2010,Feliachi_2022}. As a matter of fact, the long-range nature of the
interactions has important consequences in the structure of different physical systems such as plasmas~\cite{Nicholson_1992,Kiessling_2003},
two-dimensional fluids~\cite{Onsager_1949,Miller_1990,Robert_1991,Chavanis_2002_b,Bouchet_2009,Venaille_2012}, systems with wave-particle
interactions~\cite{Barre_2004,Barre_2005} and self-gravitating
systems~\cite{Antonov_1962,Lynden-Bell_1968,Thirring_1970,Padmanabhan_1990,Lynden-Bell_1999,Chavanis_2002,Chavanis_2006}.

Since long-range interactions couple the constituents of the system over large distances, systems with these interactions are intrinsically
non-additive~\cite{Campa_2014,Campa_2009}. This fact leads to the possibility of ensemble inequivalence~\cite{Thirring_1970,Padmanabhan_1990,Lynden-Bell_1999,Chavanis_2002,Chavanis_2006,Ellis_2000,Barre_2001,Bouchet_2005},
so equilibrium configurations strongly depend on the particular
constraints, defined by the set of control parameters, imposed on the system. In addition, from a thermodynamic point of
view~\cite{Latella_2015}, non-additivity leads to an additional degree of freedom that modifies the usual Gibbs-Duhem
equation~\cite{Latella_2013}. For macroscopic, short-range interacting systems, this equation establishes that the chemical potential
$\mu$, temperature $T$ and pressure $P$ are not independent and therefore cannot be taken together as a set of control parameters to
define equilibrium states. Instead, due to non-additivity, long-range interacting systems in the thermodynamic limit can reach states
of equilibrium with $\mu$, $T$ and $P$ as control parameters~\cite{Latella_2017}. The statistical ensemble for this set of control parameters
is called {\it unconstrained ensemble}, describing completely open systems in which energy, volume and number of particle fluctuate. It is
worth noting that the additional degree of freedom giving rise to the independence between $\mu$, $T$ and $P$ can also be realized in small
systems with short-range interactions~\cite{Hill_2013,Hill_2001}. This independence is lost, however, if the system approaches the
thermodynamic limit and the interactions remain short-range; the system becomes additive in this limit.

Since macroscopic systems with short-range interactions do not attain equilibrium states in completely open conditions~\cite{Frenkel},
numerical methods to simulate equilibrium properties with $\mu$, $T$ and $P$ as independent control parameters have received little
attention. Therefore, we recently proposed a Monte Carlo (MC) method for simulations in the unconstrained ensemble~\cite{Latella_2021};
this method is based on the Metropolis algorithm in a similar way as the schemes for other ensembles~\cite{Frenkel}, and it was
illustrated in~\cite{Latella_2021} for some simple non-additive systems that do not exhibit collective effects leading to phase transitions.
Here, we consider a more complex model that does exhibit phase transitions; the model is first analyzed theoretically, before applying to
it the MC scheme related to the unconstrained ensemble.

The system we examine here was introduced in its original form by Thirring~\cite{Thirring_1970} as a simplified version of a self-gravitating
gas of point-like particles, showing the possibility of negative specific heat in the microcanonical ensemble and the corresponding
inequivalence with the canonical case (see also references~\cite{Campa_2016,Trugilho_2022}). The model was later modified~\cite{Latella_2017}
to include both attractive and repulsive interactions in order to demonstrate that long-range interactions can lead to equilibrium states in
the unconstrained ensemble. By further modifying the model considering the particles as hard spheres, i.e., by assuming particles of finite
size, it was subsequently shown~\cite{Campa_2020} that first-order phase transitions may be observed under completely open conditions.
The theoretical approach employed in~\cite{Campa_2020} to include the effect of the finite size of the particles, however, was limited to the
excluded-volume approximation implemented as a working hypothesis in a manner similar to previous work on self-gravitating
systems~\cite{Padmanabhan_1990,Aronson_1972}. As is well known, the excluded-volume approximation for hard spheres is exact only in
$d=1$ spatial dimensions~\cite{Tonks_1936,Kac_1963}.

In the present work, we consider the modified Thirring model in $d=1,2,3$ dimensions with finite-size particles beyond the excluded-volume
approximation. Our theoretical framework encompasses a description of hard-core interactions in terms of an arbitrary local equation of state
as well as the global coupling induced by long-range interactions. A similar approach, with long-range interactions described
by a mean-field and short-range interactions described by a local equation of state, was employed in
references~\cite{Chavanis_2011,Chavanis_2019}. We focus on equilibrium configurations of the model in the unconstrained
ensemble and compare them with the corresponding configurations at fixed number of particles, in the isothermal-isobaric ensemble. By means
of MC simulations, we show not only that the system attains equilibrium states in the unconstrained ensemble but also that it exhibits
first-order phase transitions, in agreement with theoretical predictions. To reproduce the location of phase transitions, in the theory we
also introduce an approximate correction accounting for the finite number of particles in the simulations. We would like to underline that
the material and the results presented in this work represent the convergence of different issues related to the physics of non-additive
systems: the existence of equilibrium states in completely open conditions, theoretically analyzed with the unconstrained ensemble; the
possibility of having phase transitions in this ensemble; the numerical simulations of non-additive systems in completely open conditions with
a purposedly envisaged MC scheme. In particular, our results highlight a reach phenomenology in non-additive systems under completely open
conditions which, as noted above, cannot be realized in macroscopic systems with short-range interactions. This work sheds light on the
behaviour of systems exchanging heat, work and matter with their surroundings, a situation that has been poorly explored so far.

The paper is organized as follows. In section~\ref{sec:model}, we describe the system and define a set of reduced variables that are used
throughout the text. In section~\ref{sec:EVA}, we briefly summarize the theoretical approach in the excluded-volume approximation, while
in section~\ref{sec:beyond_EVA}, the theory is formulated for an arbitrary local equation of state accounting for the hard-core interactions.
In sections~\ref{sec:MC_unconstrained} and~\ref{sec:MC_isothermal}, we describe the MC simulation schemes in the unconstrained and
isothermal-isobaric ensembles, respectively. In section~\ref{sec:results}, we show our simulation results including a comparison with the
theoretical approach. Finally, in section~\ref{sec:discussion}, we present a discussion with concluding remarks.

\section{The modified Thirring model}
\label{sec:model}

\begin{figure}
\centering
\includegraphics[scale=0.8]{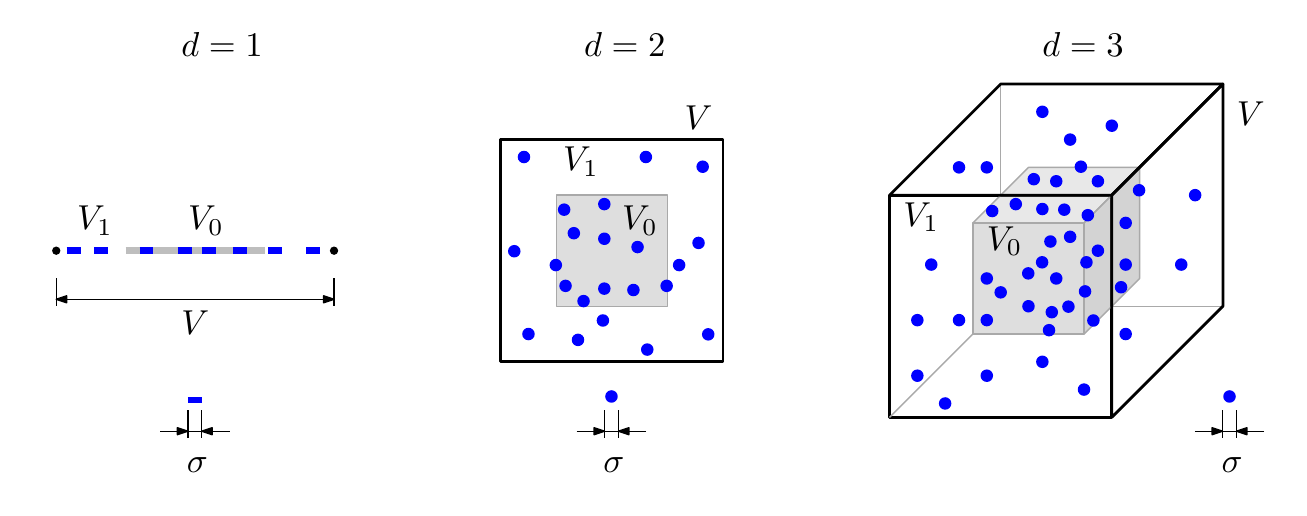}
\caption{Representation of the system for the different dimensions $d$. The system has a total $d$-dimensional volume $V$ and two internal
regions of volume $V_0$ (the core) and $V_1$ (the external part). In $d=1$ the external region consists of two disconnected portions whose
joint volume is $V_1$. The particles are rods of length $\sigma$ in $d=1$, disks of diameter $\sigma$ in $d=2$ and spheres of diameter
$\sigma$ in $d=3$.}
\label{sketch}
\end{figure}

We consider $N$ particles in a $d$-dimensional system of volume $V$, which corresponds to the length of a segment in $d=1$, the area of a
surface in $d=2$, and the usual volume of a region in $d=3$. Particles in the model are assumed to be hard spheres of diameter $\sigma$ in
$d$ dimensions which correspond to rods of length $\sigma$ in $d=1$, disks of diameter $\sigma$ in $d=2$, and spheres of diameter $\sigma$
in $d=3$, as sketched in figure~\ref{sketch}. The Hamiltonian of the system is
\begin{equation}
\mathcal{H}=\sum_{i}^N\frac{\mathbf{p}_i^2}{2m} + W(\mathbf{q}^N)
\end{equation}
with a potential energy of the form
\begin{equation}
W(\mathbf{q}^N) =\sum_{i>j}^N\phi_\mathrm{lr}(\mathbf{q}_i,\mathbf{q}_j) + \sum_{i>j}^N\phi_\mathrm{hc}(\mathbf{q}_i,\mathbf{q}_j),
\end{equation}
where $\mathbf{p}_i$ and $\mathbf{q}_i$ are the $d$-dimensional momentum and position of the center of the $i$-th particle, respectively, and
$\mathbf{q}^N\equiv(\mathbf{q}_1,\dots,\mathbf{q}_N)$. The hard-core potential is given by
\begin{equation}
\phi_\mathrm{hc}(\mathbf{q}_i,\mathbf{q}_j)=\left\{
\begin{array}{cc}
 \infty & \mbox{if } |\mathbf{q}_i-\mathbf{q}_j|<\sigma\\
 0      & \mbox{if } |\mathbf{q}_i-\mathbf{q}_j|\geq \sigma
\end{array}\right.,
\end{equation}
and the long-range potential has the form
\begin{equation}
\phi_\mathrm{lr}(\mathbf{q}_i,\mathbf{q}_j) = 
-2\nu\left[ \theta_{V_0}(\mathbf{q}_i) \theta_{V_0}(\mathbf{q}_j)
+b\theta_{V_1}(\mathbf{q}_i) \theta_{V_1}(\mathbf{q}_j)\right],
\label{interaction_potential}
\end{equation}
with $\nu>0$ and $b$ being constants. Here $V_0$ and $V_1$ are the $d$-dimensional volumes of internal regions of the system, such that
\begin{equation}
V=\sum_kV_k,\qquad k=0,1, 
\end{equation}
and the long-range interactions are defined in terms of the functions
\begin{equation}
\theta_{V_k}(\mathbf{q}_i) = \left\{
\begin{array}{cc}
 1 & \mbox{if } \mathbf{q}_i\in V_k\\
 0 & \mbox{if } \mathbf{q}_i\notin V_k
\end{array}
\right..
\end{equation}
The volume $V_0$ corresponds to a central region (the core) which is always fixed and given by the interaction potential, while the volume
$V_1$ corresponding to the external region is allowed to fluctuate depending on the external constraints imposed on the systems.
Since the interactions specified by $\phi_\mathrm{lr}(\mathbf{q}_i,\mathbf{q}_j)$ are constant within each of these two regions, the
associated potential energy is given by
\begin{equation}
\sum_{i>j}^N\phi_\mathrm{lr}(\mathbf{q}_i,\mathbf{q}_j) = \sum_kW_k
\label{potential_energy_Thirring}
\end{equation}
with
\begin{equation}
W_0 =-\nu N_0(N_0-1),\qquad W_1 =-\nu b N_1(N_1-1),
\end{equation}
where $N_0$ and $N_1$ are the number of particles in $V_0$ and in $V_1$ for a given configuration, respectively, in such a way that 
\begin{equation}
N=\sum_kN_k. 
\end{equation} 
We note that Thirring~\cite{Thirring_1970} considered the case of point-like particles, $\sigma=0$, in $d=3$ with attractive interactions
only, $b=0$. The case $b<0$ leading to repulsive interactions was considered in~\cite{Latella_2017}, whereas $b<0$ and $\sigma\neq0$ was
investigated in~\cite{Campa_2020}.

\subsection{The unconstrained ensemble and the replica energy}

We are interested in describing equilibrium configurations of completely open systems in the unconstrained ensemble, in which the control
parameters are the chemical potential $\mu$, pressure $P$ and temperature $T$. Here the pressure $P$ corresponds to the force acting on the
end points of the system in $d=1$, to the force per unit length acting on the boundary of the system in $d=2$, and to the force per unit
surface applied to the walls of the container in $d=3$. The unconstrained partition function is given by~\cite{Hill_2013,Latella_2017}
\begin{equation}
 \Upsilon(\mu,P,T)=\beta P\sum_{N}e^{\beta\mu N}\int \mathrm{d}V\, e^{-\beta P V} Z(N,V,T)
\label{unconstrained_partition_function}
\end{equation}
where $\beta=1/T$ (we use units in which $k_B=1$), we have included the factor $\beta P$ to make the partition function
dimensionless~\cite{Hill_Statistical_Mechanics}, and introduced the canonical partition function
\begin{equation}
Z(N,V,T)= \int \frac{\mathrm{d}^{dN}\mathbf{q}}{\lambda_T^{dN}N!} \ e^{-\beta W(\mathbf{q}^N)}
\label{canonical_partition_function}
\end{equation}
in which $\mathrm{d}^{dN}\mathbf{q}\equiv \mathrm{d}^{d}\mathbf{q}_1\dots \mathrm{d}^{d}\mathbf{q}_N$ and $\lambda_T$ is the thermal
de Broglie wavelength.  In the unconstrained ensemble, the thermodynamics of the system is derived from the appropriate free
energy~\cite{Hill_2013,Latella_2017} $\mathscr{E}=-T\ln\Upsilon(\mu,P,T)$ which is called \textit{replica energy}.

It is known that in long-range systems the non-additivity of the energy does not allow the standard derivation of the canonical ensemble from
the microcanonical ensemble as it is usually done for additive systems~\cite{Campa_2014,Campa_2009}. However, the use of the canonical ensemble for systems with long-range interactions can be justified in various ways. For instance, for a long-range system one has to envisage the exchange of heat with the surroundings made of different components, that interact with the particles of the system via a short-range interaction~\cite{Baldovin_1,Baldovin_2}. In this case, although the particles of the system have a long-range interaction among themselves, the total energy of the system plus the bath is still the sum of the energy of the system and the energy of the bath. Another approach is the one of coupling each particle of the long-range system to stochastic noise, which simulates a bath at fixed temperature~\cite{Chavanis_2006_b}.

We will also compare the equilibrium states in the unconstrained ensemble with those in the isothermal-isobaric ensemble. In the latter,
the control parameters are $N$, $P$ and $T$ and the corresponding partition function reads
\begin{equation}
 \Delta(N,P,T)=\beta P\int \mathrm{d}V\, e^{-\beta P V} Z(N,V,T).
\label{isobaric_partition_function}
\end{equation}
In this ensemble, the thermodynamics of the system follows from the Gibbs free energy $G(N,P,T)=-T\ln\Delta(N,P,T)$.
Notice that the unconstrained and isothermal-isobaric partitions functions are related through
$\Upsilon(\mu,P,T)=\sum_{N}e^{\beta\mu N}\Delta(N,P,T)$.

A typical feature of long-range interacting systems is that the potential energy scales as $N^2$. Thus, interesting phenomena such as
phase transitions commonly occur when the temperature is of order $N$, in a way that kinetic and potential energies are of the same order
in equilibrium states. When the temperature is a control parameter, it can be suitably chosen, of order $N$, to observe such phenomena.
Furthermore, if the physical constraints on the system corresponds to an ensemble in which $V$ or $N$ can be controlled at fixed $T$, the
size of the system is defined by these parameters and a thermodynamic limit can be taken sending them to infinity keeping the appropriate
ratio finite. In the unconstrained ensemble, however, neither $V$ nor $N$ are control parameters and, therefore, this way of taking the
thermodynamic limit is not applicable. Exploiting the fact that the size of the system depends on variables such as $T$, $P$ and $\mu$
when the system is non-additive (this variables are intensive in additive systems), a proper thermodynamic limit can be taken by
choosing $T$, $P$ or $\mu$ in a manner that the system becomes macroscopic, keeping, for instance, the average particle density finite.
As we show below with theory and MC simulations, the macroscopic limit for the Thirring model in the unconstrained ensemble is achieved
as $T/\nu\to \infty$, so the average number of particles $\bar{N}$ diverges in this limit,
$\bar{N}\to\infty$. 

\subsection{Reduced variables}

In order to study some features of the model, for convenience we define a set of dimensionless variables that we will use throughout the
paper. We introduce the reduced volume $v$, reduced number of particles $x_k$ in region $k$, and reduced total number of particles $x$
which are given by
\begin{equation}
v=\frac{V-V_0}{V_0},\qquad x_0=\frac{\nu N_0}{T},\qquad x_1=\frac{\nu N_1}{T},\qquad x=x_0+x_1.
\label{reduced_variables_1}
\end{equation}
We also introduce the exclusion parameter $a$, reduced pressure $p$ and reduced chemical potential $\xi$ which are defined as
\begin{equation}
a=\frac{TB}{\nu V_0},\qquad p=\frac{\nu V_0}{T^2}P, \qquad \xi =\frac{\mu_T-\mu}{T}, 
\label{reduced_variables_2}
\end{equation}
where
\begin{equation}
\mu_T=T\ln\left(\frac{T\lambda_T^d}{\nu V_0}\right)
\end{equation}
and 
\begin{equation}
B=2^{d-1} \Omega_d\sigma^d, \qquad \Omega_d=\frac{(\pi/4)^{d/2}}{\Gamma(1+{\textstyle\frac{d}{2}})} .
\end{equation}
Here $\Omega_d$ is the $d$-dimensional volume of a sphere of unit diameter, $\Gamma(x)$ being the gamma function. The parameter $B$ is the
usual second virial coefficient of the hard-sphere fluid (without the long-range interactions) and accounts for the $d$-dimensional
excluded volume per particle. The above set of reduced variables generalize to the case of arbitrary dimension $d$ the reduced variables
defined in Ref.~\cite{Campa_2020}.

\section{Excluded-volume approximation}
\label{sec:EVA}
 
Hard-core interactions in this model were considered in Ref.~\cite{Campa_2020} using the excluded-volume approximation to describe
equilibrium states in the different ensembles. For the sake of clarity, in this section we summarize this approach in the unconstrained
ensemble for arbitrary dimension $d$. This theoretical framework is extended beyond the excluded-volume approximation in
section~\ref{sec:beyond_EVA}.

In order to account for the hard-core interactions, here we consider excluded-volume effects characterized by the $d$-dimensional excluded
volume per particle $B$. In this approximation, the canonical partition function $Z_k$ at temperature $T$ of each subsystem in region $k=0,1$
is given by
\begin{equation}
Z_k(N_k,V_k,T)=\frac{\left(V_k-N_kB\right)^{N_k}e^{-\beta W_k}}{\lambda_T^{dN_k}N_k!}.
\end{equation}
Furthermore, the two internal regions are allowed to exchange particles, so that the total partition function of the system reads
\begin{equation}
Z(N,V,T)=\sum_{N_0,N_1}\delta_{N,N_0+N_1}Z_0(N_0,V_0,T)Z_1(N_1,V_1,T),
\label{canonical_approximation}
\end{equation}
where the Kronecker delta fixes the total number of particles to $N$ and the total volume is fixed as $V=\sum_kV_k$. The above partition
function neglects hard-core interactions that take place at the boundary between the subsystems. 

In the unconstrained ensemble, the control parameters are $\mu$, $P$ and $T$. The unconstrained partition function in this approximation
can be obtained by using the canonical partition function (\ref{canonical_approximation}) in expression
(\ref{unconstrained_partition_function}). Following the steps in~\cite{Campa_2020}, one arrives at
\begin{equation}
\Upsilon(\mu,P,T)= \int \mathrm{d}V_1\sum_{N_0,N_1} e^{-\beta \hat{\mathscr{E}}(V_1,N_0,N_1)}
\label{upsilon}
\end{equation}
with
\begin{equation}
\hat{\mathscr{E}}(V_1,N_0,N_1) = \sum_k\left[PV_k  +W_k + T N_k\left( \frac{d}{2} -s_k-\frac{\mu}{T}\right)\right],
\label{replica_functional}
\end{equation}
where we have made explicit the contribution of the local entropy per particle
\begin{equation}
s_k= -\ln \left(\frac{N_k }{V_k}\lambda_T^d\right) +\frac{d+2}{2} -\ln \left(\frac{V_k}{V_k-N_kB}\right)
\label{local_entropy_excluded}
\end{equation}
in the excluded-volume approximation.
Notice that since $V_0$ is fixed by the interaction potential, we have replaced the integration over $V$ by an integration over $V_1$ in
equation (\ref{upsilon}). The replica energy describing equilibrium states in the unconstrained ensemble can be obtained by
computing (\ref{upsilon}) in a saddle-point approximation, so that
\begin{equation}
\mathscr{E}=\min_{\{V_1,N_0,N_1\}} \hat{\mathscr{E}}(V_1,N_0,N_1).
\label{min_replica_excluded}
\end{equation}

The saddle-point equations minimizing $\hat{\mathscr{E}}$ in the $d$-dimensional case follow as in~\cite{Campa_2020}. 
In terms of the reduced variables (\ref{reduced_variables_1}) and (\ref{reduced_variables_2}), the reduced replica energy
$\hat{\varphi}_u= \nu \hat{\mathscr{E}}/T^2$ can be written as
\begin{eqnarray}
\hat{\varphi}_u(v,x_0,x_1) 
&= x_0\left[ \ln \left(\frac{x_0}{1-ax_0}\right)-1 \right]
+ x_1\left[\ln \left(\frac{x_1}{v-ax_1}\right)-1\right] \nonumber\\
&+p(v+1) +\xi(x_0+x_1) -x_0^2  -bx_1^2,
\label{replica_functional_reduced}
\end{eqnarray}
and the minimization problem (\ref{min_replica_excluded}) becomes
\begin{equation}
\varphi_u=\min_{\{v,x_0,x_1\}} \hat{\varphi}_u(v,x_0,x_1) ,
\label{min_replica_excluded_reduced}
\end{equation}
where $\varphi_u=\nu\mathscr{E}/T^2$. Derivatives with respect to $v$, $x_0$ and $x_1$ lead to
\begin{eqnarray}
\bar{v} &=& \frac{1+ap}{2b p} \left(\ln p  + a p +\xi\right),\label{eq_1}\\
\bar{x}_0 &=&
\frac{1}{2}\ln\left(\frac{\bar{x}_0 }{1-a\bar{x}_0} \right) + \frac{a\bar{x}_0}{2(1-a\bar{x}_0)} + \frac{\xi}{2}, \label{eq_2}\\
\bar{x}_1 &=& 
\frac{1}{2b}\left(\ln p + a p +\xi \right),\label{eq_3}
\end{eqnarray}
where the bar over a given quantity indicates that the quantity is solution of the minimization problem and represents an average value.
These are the same equations obtained in~\cite{Campa_2020}. We highlight that using the reduced variables $a$, $p$ and $\xi$, the
saddle-point equations in the excluded-volume approximation are invariant under change of dimension $d$. Notice also that these equations
do not depend explicitly on the temperature, so $T$ is just a scaling factor here. Thus, within the excluded-volume approximation, phase
transitions and critical points of the model studied in~\cite{Campa_2020} correspond to any dimension $d$. However, as noted previously,
this approximation is exact for $d=1$, while for $d>1$ is only valid at low densities. Thus, while the approach in~\cite{Campa_2020} is
accurate when restricting the system to the 1-dimensional case, it fails to quantitatively describe states of relatively high densities
for $d>1$. We nevertheless emphasize that interesting features of the model such that ensemble inequivalence and the realization of phase
transitions in the unconstrained ensemble, which are qualitatively described with the excluded-volume approximation for $d>1$, remain
valid with the more accurate description that we provide below.

\section{Beyond the excluded-volume approximation}
\label{sec:beyond_EVA}

Here we improve the description of the system given in the previous section by considering an arbitrary equation of state for the
$d$-dimensional hard-sphere fluid. This allows us to go beyond the excluded-volume approximation by selecting an equation of state
appropriate for the concrete dimension $d$ of the system under examination.

The basic idea in this approach is that the local entropy per particle is characterized by short-range interactions only, so locally it
satisfies the usual thermodynamic relations for short-range systems~\cite{Latella_2013,Latella_2015,Chavanis_2020} with all thermodynamic
quantities properly defined at the local level. Thus, consider the local internal energy per particle $u_k=\frac{d}{2}T_k$ and local
specific volume $v_k=V_k/N_k$ in region $k=0,1$ for a $d$-dimensional gas of hard spheres, where $T_k$ is the local temperature. The local
entropy per particle $s_k=s_k(u_k,v_k)$ in these regions of the system can be written as
\begin{equation}
s_k= s^\mathrm{id}_k + s^\mathrm{ex}_k,
\label{local_entropy}
\end{equation}
where
\begin{equation}
s^\mathrm{id}_k= -\ln\left(\frac{N_k}{V_k}\lambda_T^d\right) +\frac{2+d}{2}
= \ln\left(cv_k u_k^{d/2}\right) +\frac{2+d}{2}
\label{local_ideal_entropy}
\end{equation}
is the local entropy per particle of an ideal gas and $s^\mathrm{ex}$ accounts for the excess entropy of the hard spheres, with $c$ being
a constant. The entropy locally satisfies the thermodynamic relations~\cite{Latella_2013}
\begin{eqnarray}
\frac{1}{T_k}=\left(\frac{\partial s_k}{\partial u_k}\right)_{v_k},\label{local_temperature}\\ 
\frac{p_k}{T_k}=\left(\frac{\partial s_k}{\partial v_k}\right)_{u_k}\label{local_pressure},
\end{eqnarray}
where $p_k$ is the local pressure. (The local specific volume $v_k$ and the local pressure $p_k$ should not be confused with the
dimensionless reduced volume $v$ and pressure $p$.) If an equation of state for the local pressure is given through the compressibily
factor $f(\eta_k)$, namely,
\begin{equation}
\label{gen_eq_state}
f(\eta_k)=\frac{p_k v_k}{T_k},
\end{equation}
equations (\ref{local_temperature}) and (\ref{local_pressure}) are satisfied for an excess entropy given by
\begin{equation}
s^\mathrm{ex}_k(\eta_k)= -\int_0^{\eta_k}\frac{f(\eta)-1}{\eta}\mathrm{d}\eta,
\label{local_excess_entropy}
\end{equation}
where $\eta_k=\Omega_d\sigma^d/v_k$ is the local packing fraction. We emphasize that it is possible to write the equation of state in the form
(\ref{gen_eq_state}), with the left hand side depending only on the packing fraction (i.e., on the density) and independent of the temperature,
since we are considering hard spheres, for which the virial coefficients do not depend on the temperature.

The excluded volume approximation of the previous section, for instance, is obtained by considering
\begin{equation}
f(\eta_k)=\frac{V_k}{V_k-B N_k}=\frac{1}{1-2^{d-1}\eta_k},
\end{equation}
which is exact only for $d=1$.
For a more accurate treatment of the $d$-dimensional case, here we consider the exact equation of state~\cite{Tonks_1936}
\begin{equation}
f(\eta)= \frac{1}{1-\eta}, \qquad d=1,
\label{eos_d1}
\end{equation}
the Henderson~\cite{Henderson_1975} equation of state
\begin{equation}
f(\eta)=\frac{1+\frac{1}{8}\eta^2}{(1-\eta)^2},\qquad d=2,
\label{eos_d2}
\end{equation}
and the Carnahan-Starling~\cite{Carnahan_1969} equation of state
\begin{equation}
f(\eta)=\frac{1+\eta+\eta^2-\eta^3}{(1-\eta)^3}, \qquad d=3.
\label{eos_d3}
\end{equation}
While more accurate equations of state have been proposed (see~\cite{Santos_1995,Robles_2014}, for instance), the accuracy of those
considered here is enough for our purpose because packing fractions in our examples are always below the freezing point.
With these $f(\eta)$, the integrals appearing in equation (\ref{local_excess_entropy}) are given by
\begin{equation}
\int_0^{\eta}\frac{f(y)-1}{y}\mathrm{d}y= -\ln(1-\eta),\qquad d=1,
\end{equation}
\begin{equation}
\int_0^{\eta}\frac{f(y)-1}{y}\mathrm{d}y= \frac{9\eta}{8(1-\eta)}-\frac{7}{8}\ln(1-\eta),\qquad d=2,
\end{equation}
\begin{equation}
\int_0^{\eta}\frac{f(y)-1}{y}\mathrm{d}y= \frac{4\eta-3\eta^2}{(1-\eta)^2},\qquad d=3.
\end{equation}

Assuming thermodynamic equilibrium, the local temperature is the same in the two regions, $T=T_k$, and the pressure in region $k=1$
corresponds to the pressure imposed on the boundary of the system, $P=p_1$. Moreover, the chemical potential takes the same value in the
two regions: in the unconstrained ensemble this is achieved by imposing equilibrium with an external reservoir with fixed $\mu$, while in
the isothermal-isobaric this condition follows from internal equilibrium with fixed $N$. Because of the long-range interactions, the
packing fractions $\eta_0$ and $\eta_1$ as well as the number of particles $N_0$ and $N_1$ in the two regions are different. Including
these interactions, the total entropy and energy of the system for a given configuration are
\begin{eqnarray}
S&= \sum_kN_ks_k,\\
E&= \frac{d}{2}T\sum_k N_k + \sum_kW_k,
\end{eqnarray}
respectively. Below we specify the equilibrium conditions in both the unconstrained and isothermal-isobaric ensembles.

\subsection{Unconstrained ensemble}

The replica energy $\hat{\mathscr{E}}=E-TS+PV-\mu N$ can be written as
\begin{equation}
\hat{\mathscr{E}}(V_1,N_0,N_1) =\sum_k\left[PV_k + W_k + T N_k \left( \frac{d}{2} -s_k -\frac{\mu}{T} \right)\right]
\end{equation}
for arbitrary $V_1$, $N_0$ and $N_1$.
This expression is the same as in equation (\ref{replica_functional}), but now the local entropy is given by (\ref{local_entropy}) with the
excess entropy (\ref{local_excess_entropy}). The replica energy of equilibrium configurations is then given by
\begin{equation}
\mathscr{E}=\min_{\{V_1,N_0,N_1\}} \hat{\mathscr{E}}(V_1,N_0,N_1),
\label{min_replica_CS}
\end{equation}
which is a function of $\mu$, $P$ and $T$.
In terms of the reduced variables (\ref{reduced_variables_1}), (\ref{reduced_variables_2}) and taking $\varphi_u= \nu \mathscr{E}/T^2$ and
$\hat{\varphi}_u= \nu \hat{\mathscr{E}}/T^2$, this variational problem can be stated as
\begin{equation}
\varphi_u=\min_{\{v,x_0,x_1\}} \hat{\varphi}_u(v,x_0,x_1) ,
\label{min_replica_excluded_reduced_2}
\end{equation}
where 
\begin{eqnarray}
\hat{\varphi}_u (v,x_0,x_1)
&= x_0\left[\ln\left(x_0\right)  +\int_0^{\eta_0}\frac{f(\eta)-1}{\eta}\mathrm{d}\eta -1\right] \nonumber\\
&+ x_1\left[\ln\left(\frac{x_1}{v}\right)   +\int_0^{\eta_1}\frac{f(\eta)-1}{\eta}\mathrm{d}\eta -1\right] \\
&+ p(v+1) +\xi(x_0+x_1)- x_0^2 -bx_1^2\nonumber
\label{replica_energy_reduced}
\end{eqnarray}
and we have used that $N_k(N_k-1)\approx N_k^2$, meaning that we neglect here correction terms $\nu/T$ that vanish in the limit
$T/\nu\to\infty$. In fact, as we have remarked in section \ref{sec:model}, the macroscopic limit for our model in the unconstrained
ensemble is achieved with $T/\nu \to \infty$. We still have to show explicitly this fact, which is done below, in section
\ref{sec:finite_temp_correction}, when dealing with finite size corrections. Using the reduced variables, the packing fractions read
\begin{eqnarray}
\eta_0 &= \frac{a}{2^{d-1}} x_0, \label{eta_0}\\
\eta_1 &= \frac{a}{2^{d-1}} \frac{x_1}{v}\label{eta_1}.
\end{eqnarray}
Setting to zero the derivatives of $\hat{\varphi}_u$ with respect to $v$, $x_0$ and $x_1$ leads to
\begin{eqnarray}
\frac{2^{d-1}p}{a\bar{\eta}_1}&= f(\bar{\eta}_1),\label{EOS_reduced}\\
2\bar{x}_0
&=
\ln\left(\bar{x}_0\right) 
+\int_0^{\bar{\eta}_0}\frac{f(\eta)-1}{\eta}\mathrm{d}\eta 
+f(\bar{\eta}_0)-1 +\xi,\label{x_0}\\
2b\bar{x}_1
&=
\ln\left(  \frac{2^{d-1}}{a} \bar{\eta}_1 \right) 
+\int_0^{\bar{\eta}_1}\frac{f(\eta)-1}{\eta}\mathrm{d}\eta 
+f(\bar{\eta}_1) -1 +\xi,\label{x_1}
\end{eqnarray}
with $\bar{\eta}_k$ being the equilibrium packing fraction in region $k$ which are given by
\begin{eqnarray}
\bar{\eta}_0 &= \frac{a}{2^{d-1}} \bar{x}_0, \label{red_eta_0}\\
\bar{\eta}_1 &= \frac{a}{2^{d-1}} \frac{\bar{x}_1}{\bar{v}}\label{red_eta_1}.
\end{eqnarray}

The solution to the system of equations (\ref{EOS_reduced})-(\ref{red_eta_1}) can be obtained as follows. Equation (\ref{EOS_reduced}) is
first solved for $\bar{\eta}_1$, which can have at most one real solution in the range $0<\bar{\eta}_1<1$ for a physically consistent equation
of state $f(\bar{\eta}_1)$. Then $\bar{\eta}_1$ is replaced in (\ref{x_1}) yielding $\bar{x}_1$, and $\bar{v}$ is obtained from
(\ref{red_eta_1}). Since $\bar{v}$ and $\bar{x}_1$ are uniquely determined by the control parameters, the occurrence of phase transitions,
if any, is related to the existence of multiple solutions for $\bar{x}_0$ in equation (\ref{x_0}). If multiple solutions exist, the
equilibrium state is characterized by $\bar{x}_0$ minimizing the replica energy. As in~\cite{Campa_2020}, this fact allows us to study the
occurrence of phase transitions by considering only the terms in $\hat{\varphi}_u(\bar{v},x_0,\bar{x}_1)$ that depend on $x_0$, that is
\begin{equation}
\widetilde{\varphi}_u(x_0)= x_0\left[\ln\left(x_0\right) +\int_0^{\eta_0}\frac{f(\eta)-1}{\eta}\mathrm{d}\eta +\xi -1\right]-x_0^2,
\label{replica_energy_reduced_x0}
\end{equation}
which here is written explicitly in terms of the equation of state $f(\eta)$. Then the equilibrium value $\bar{x}_0$ can also be obtained by
minimizing the free energy $\widetilde{\varphi}_u(x_0)$.

\subsection{Isothermal-isobaric ensemble}
\label{theory_iso_ensemble}

The Gibbs free energy $\hat{G}=E-TS+PV$ for arbitrary $V_1$ and $N_0$ can be written as
\begin{equation}
\hat{G}(V_1,N_0) =\sum_k\left[PV_k + W_k + T N_k \left( \frac{d}{2} -s_k \right)\right],
\label{hat_Gibbs}
\end{equation}
where the total number of particles $N$ is fixed in such a way that $N_1=N-N_0$ and the local entropy per particle is given by
(\ref{local_entropy}) with the excess entropy (\ref{local_excess_entropy}). The free energy of equilibrium configurations follows from
\begin{equation}
G=\min_{\{V_1,N_0\}} \hat{G}(V_1,N_0),
\label{min_Gibbs}
\end{equation}
which is a function of $N$, $P$ and $T$.
In terms of the reduced variables (\ref{reduced_variables_1}) and (\ref{reduced_variables_2}), this variational problem can be stated as
\begin{equation}
\varphi_i=\min_{\{v,x_0\}} \hat{\varphi}_i(v,x_0) ,
\label{min_Gibbs_reduced}
\end{equation}
where $\varphi_i= \nu G/T^2$ and $\hat{\varphi}_i= \nu \hat{G}/T^2$ is expressed as
\begin{eqnarray}
\hat{\varphi}_i (v,x_0)
&= x_0\left[\ln\left(x_0\right)  +\int_0^{\eta_0}\frac{f(\eta)-1}{\eta}\mathrm{d}\eta \right] \nonumber\\
&+ (x-x_0)\left[\ln\left(\frac{x-x_0}{v}\right)   +\int_0^{\eta_1}\frac{f(\eta)-1}{\eta}\mathrm{d}\eta \right] \label{Gibbs_reduced}\\
&+ p(v+1) - x_0^2 -b(x-x_0)^2 +x\left(\frac{\mu_T}{T}-1\right)\nonumber
\end{eqnarray}
with $\eta_0$ given by (\ref{eta_0}) and
\begin{equation}
\eta_1 = \frac{a}{2^{d-1}} \frac{x-x_0}{v}. 
\label{eta_1_iso}
\end{equation}
Setting to zero the derivatives of $\hat{\varphi}_i$ with respect to $v$ and $x_0$ leads to
\begin{eqnarray}
\frac{ap}{2^{d-1}\bar{\eta}_1} &=  f(\bar{\eta}_1) ,\label{EOS_reduced_iso}\\
2\bar{x}_0
&= \ln\left(\bar{x}_0\right)  +\int_0^{\bar{\eta}_0}\frac{f(\eta)-1}{\eta}\mathrm{d}\eta+f(\bar{\eta}_0)
+2b(x-\bar{x}_0) \nonumber\\
&- \ln\left(\frac{2^{d-1}}{a}\bar{\eta}_1\right)   -\int_0^{\bar{\eta}_1}\frac{f(\eta)-1}{\eta}\mathrm{d}\eta 
-f(\bar{\eta}_1),\label{x_0_iso}
\end{eqnarray}
where 
\begin{eqnarray}
\bar{\eta}_0 &= \frac{a}{2^{d-1}} \bar{x}_0, \label{red_eta_0_iso}\\
\bar{\eta}_1 &= \frac{a}{2^{d-1}} \frac{x-\bar{x}_0}{\bar{v}}\label{red_eta_1_iso}.
\end{eqnarray}

Equilibrium states in the isothermal-isobaric ensemble are characterized by the solution to the system of equations
(\ref{EOS_reduced_iso})-(\ref{red_eta_1_iso}).  The packing fraction $\bar{\eta}_1$ is directly obtained from equation (\ref{EOS_reduced_iso}),
which is replaced in (\ref{x_0_iso}) yielding $\bar{x}_0$.  With this, one has $\bar{x}_1=x-\bar{x}_0$ and $\bar{v}$ is obtained from
(\ref{red_eta_1_iso}). Analogous to what happens in the unconstrained ensemble, the occurrence of phase transitions is related to the
existence of multiple solutions for $\bar{x}_0$ in equation (\ref{x_0_iso}). If multiple solutions exist, the equilibrium state corresponds
to $\bar{x}_0$ minimizing the Gibbs free energy. By changing variables from $v$ to $\eta_1$ through equation (\ref{eta_1_iso}) and
considering only the terms in $\hat{\varphi}_i(\bar{\eta}_1,x_0)$ that depend on $x_0$, we have
\begin{eqnarray}
\widetilde{\varphi}_i (x_0)
&= x_0\left[\ln\left(x_0\right)  +\int_0^{\eta_0}\frac{f(\eta)-1}{\eta}\mathrm{d}\eta +h(\bar{\eta}_1)\right] 
- x_0^2 -b(x-x_0)^2, 
\label{Gibbs_reduced_x0}
\end{eqnarray}
where
\begin{equation}
h(\bar{\eta}_1)=-\ln\left(\frac{2^{d-1}}{a}\bar{\eta}_1 \right)   -\int_0^{\bar{\eta}_1}\frac{f(\eta)-1}{\eta}\mathrm{d}\eta -f(\bar{\eta}_1).
\label{aux_func_h}
\end{equation}
Thus, $\bar{x}_0$ can be alternatively obtained from the minimization of $\widetilde{\varphi}_i (x_0)$.

\begin{figure}
\centering
\includegraphics[scale=1]{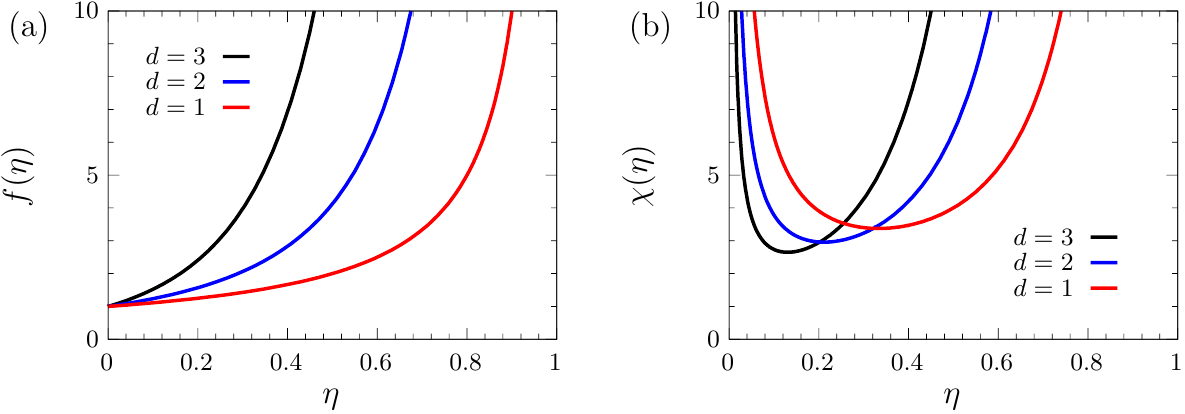}
\caption{Hard-sphere equation of state $f(\eta)$ and the function $\chi(\eta)$ for the different dimensionalities.}
\label{eos}
\end{figure}

Furthermore, in the isothermal-isobaric ensemble the chemical potential of the system is given by
\begin{equation}
\mu=\left(\frac{\partial G}{\partial N}\right)_{T,P}= \left.\frac{\partial \hat{G}}{\partial N}\right|_{\bar{V}_1,\bar{N}_0},
\end{equation}
where the expression on the right must be evaluated at $\bar{V}_1$ and $\bar{N}_0$ that minimize $\hat{G}$. Taking $N_1=N-N_0$ in the
free energy (\ref{hat_Gibbs}), we get
\begin{equation}
\mu = -2\nu b \bar{N}_1 + T \left[  \ln\left(\frac{\bar{N}_1}{\bar{V}_1}\lambda_T^d\right) +
\int_0^{\bar{\eta}_1}\frac{f(\eta)-1}{\eta}\mathrm{d}\eta +  f(\bar{\eta}_1) -1\right],
\end{equation}
where $\bar{N}_1=N-\bar{N}_0$ and $\bar{\eta}_1=\Omega_d\sigma^d \bar{N}_1/\bar{V}_1$. In terms of the reduced variables, the reduced
chemical potential takes the form
\begin{equation}
\xi= 2 b (x-\bar{x}_0) -   \ln\left(\frac{2^{d-1}}{a}\bar{\eta}_1\right) -
\int_0^{\bar{\eta}_1}\frac{f(\eta)-1}{\eta}\mathrm{d}\eta -  f(\bar{\eta}_1) +1.
\label{xi_iso}
\end{equation}
The above expression for $\xi$ is useful to compare results in the isothermal-isobaric ensemble with those obtained in the unconstrained
ensemble where $\xi$ is a control parameter. In particular, we are interested in the relation between $\xi$ and the reduced total number
of particles $x$ which are conjugated variables. At this point it is convenient to remark that the response functions $M_{T,P}$, given by
\begin{equation}
\frac{1}{M_{T,P}}=\left(\frac{\partial\mu}{\partial N}\right)_{T,P}=\left(\frac{\partial^2 G}{\partial N^2}\right)_{T,P},
\end{equation}
is not restricted to be a positive quantity in the isothermal-isobaric ensemble~\cite{Campa_2018}, since $N$ is a control parameter which
is kept fixed in equilibrium configurations. With the reduced variables, the response function can be expressed as
\begin{equation}
\frac{1}{M_{T,P}}= -\nu\frac{\partial \xi}{\partial x}
\end{equation}
at constant $T$ and $P$, so that the exclusion parameter $a$ and the reduced pressure $p$ are held constant as well. Next we argue that
$M_{T,P}$ can be negative (i.e. $\partial \xi/\partial x>0$) for the considered model, as we show in the following sections with explicit
examples.

According to equation (\ref{EOS_reduced_iso}), the packing fraction $\bar{\eta}_1$ does not depend on $x$ and thus from (\ref{xi_iso}) we have
\begin{eqnarray}
\frac{\partial \xi}{\partial x}&= 2 b -2b\frac{\partial \bar{x}_0}{\partial x}.
\end{eqnarray}
In addition, computing the derivative of equation (\ref{x_0_iso}) with respect to $x$ and rearranging terms one obtains
\begin{eqnarray}
\frac{\partial \bar{x}_0}{\partial x} =\frac{b}{1-a\chi(\bar{\eta}_0) +b } 
\end{eqnarray}
and therefore
\begin{eqnarray}
\frac{\partial \xi}{\partial x}&= 2 \left(\frac{1}{b} +\frac{1}{1-a\chi(\bar{\eta}_0)} \right)^{-1},
\end{eqnarray}
where
\begin{equation}
\chi (\eta)= \frac{ 1}{2^{d}}
\left(\frac{f(\eta)}{\eta}+\frac{\partial f(\eta)}{\partial \eta}\right) 
\end{equation}
and we have used that $\bar{\eta}_0/\bar{x}_0=a/2^{d-1}$. We note that the function $\chi(\eta)$ depends on the dimension through the
equation of state $f(\eta)$ but it does not depend on the control parameters used to specify the state of the system (the value of
$\bar{\eta}_0$ in a given configuration does depend on these parameters). The considered equations of state $f(\eta)$, given by
equations (\ref{eos_d1})-(\ref{eos_d3}), and the corresponding functions $\chi(\eta)$ are shown in figure~\ref{eos} as a function of the
packing fraction for the different dimensions. For $\eta$ small enough, $\chi(\eta)$ is relatively large and positive, so $1-a\chi$ is
negative and the slope of the curve $\xi(x)$ is always negative for $b<0$.  In the isothermal-isobaric ensemble, where $x$ is a control
parameter, we observe that $\partial \xi/\partial x$ can be positive at packing fractions for which $(1-a\chi)^{-1}>-b^{-1}$. As we see
in figure~\ref{eos}(b), the drop in $\chi(\eta)$ as $\eta$ increases is stronger when the dimension $d$ is also increased. Thus,
increasing $d$ favors the appearance of a positive slope $\partial \xi/\partial x>0$. Moreover, the value of the absolute minimum
of $\chi(\eta)$ decreases as $d$ increases, so $\partial \xi/\partial x>0$ could be observed for $d=3$, for instance, but not for $d<3$.

\subsection{Finite-size correction}
\label{sec:finite_temp_correction}

To compare theory with simulations, as we do in section~\ref{sec:results}, we have to take into account corrections that arise because
the simulations are performed with finite number of particles. We show below that these corrections are related to size effects introduced
by the fact that there is no physical barrier between the two regions of the system.

Since the spheres have a finite diameter $\sigma$, there can be particles whose center lie in one region but part of their volume reside
in the other region. Thus, the effective volume occupied by $N_k$ particles in region $k$ is somewhat larger than $V_k$. The effective
packing fractions are then smaller than those obtained using $V_k$ as the volume. In the situations of interest, the packing fraction
$\eta_1$ is small while $V_1/V_0$ is large, so here we neglect this correction for particles in region $k=1$. The reference volume $V_0$,
however, is kept fixed in the simulations and finite size effects can be important in this region, so the packing fraction $\eta_0$ should
be corrected to account for these effects. The correction can be estimated as follows by means of geometric considerations.

\begin{figure}
\centering
\includegraphics[scale=1]{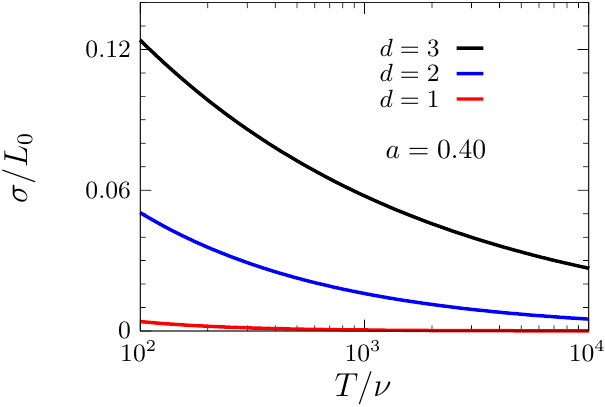}
\caption{Ratio $\sigma/L_0$ as a function of the temperature for different dimensions with an exclusion parameter $a=0.40$. For finite
temperature, $\sigma/L_0$ increases as the dimensionality increases.}
\label{sigma}
\end{figure}

Considering the finite size of the particles, the available volume for particles in region $k=0$ can be computed as
\begin{equation}
V_0'=V_0(1+\sigma/L_0)^d, 
\label{available_volume}
\end{equation}
where $L_0$ is the side of the internal region such that $V_0=L_0^d$.
Since in the simulations we fix the exclusion parameter $a$ defined in (\ref{reduced_variables_2}), we actually compute the ratio
$\sigma/L_0$ as
\begin{equation}
\frac{\sigma}{L_0}=\left(\frac{\nu a}{2^{d-1}\Omega_dT}\right)^{1/d},
\label{sigma_L0}
\end{equation}
which vanishes in the limit $T/\nu\to\infty$. For finite temperature, we have to keep this term to properly describe the available volume
(\ref{available_volume}). Moreover, the higher the dimensionality, the slower $\sigma/L_0$ approaches zero, so finite temperature effects
in the simulations are more noticeable as $d$ increases (see figure~\ref{sigma}). We remark that this behavior of $\sigma/L_0$ is due to the
use of our reduced variables, both in the theoretical treatment and in the simulations below. Given the definition of $a$ in
equations~(\ref{reduced_variables_2}), keeping $a$ fixed while $T/\nu\to\infty$ means also that the volume $V_0$ goes to infinite. Hence,
in addition to the fact that $\sigma/L_0\to0$ in this limit, the number of particles $N_0=x_0 T/\nu$ in $V_0$ can increase without limits
with fixed values of $x_0$ in equilibrium configurations. Thus, as promised above, we have shown that the thermodynamic limit is achieved
when $T/\nu \to \infty$.

To first order in $\sigma/L_0$, the difference $\Delta V_0=V_0'-V_0$ is given by $\Delta V_0\approx V_0 d \sigma/L_0 $. Furthermore,
let $\tilde{\eta}_0$ be the effective (corrected) packing fraction in $V_0$ and assume that the packing fraction in
the volume $\Delta V_0$ is the average between $\tilde{\eta}_0$ and $\eta_1$. Since there are $N_0$ particles in the total available
volume $V_0'$, we have 
\begin{equation}
\Omega_d\sigma^dN_0= \tilde{\eta}_0 V_0+\frac{1}{2}(\tilde{\eta}_0 +\eta_1)\Delta V_0.
\end{equation}
For $\eta_1/\tilde{\eta}_0\ll 1$ and to first order in $\sigma/L_0$, we are led to
\begin{equation}
\tilde{\eta}_0 \approx  \Omega_d\sigma^d\frac{N_0}{V_0}\left(1-\frac{d}{2}\frac{\sigma}{L_0}\right)=\eta_0\gamma_0,
\end{equation}
where $\eta_0$ is given by (\ref{eta_0}) in terms of the reduced variables; then the correction due to finite temperature reads
\begin{equation}
\gamma_0= 1-\frac{d}{2}\frac{\sigma}{L_0}.
\end{equation}
In this approximation, we solve the minimization problem and obtain $\bar{x}_0$ by replacing
\begin{equation}
\eta_0\to \tilde{\eta}_0 =\eta_0\gamma_0
\label{corrected_eta0}
\end{equation}
in the free energies (\ref{replica_energy_reduced_x0}) and (\ref{Gibbs_reduced_x0}) in the unconstrained and isothermal-isobaric
ensembles, respectively. As we show in section~\ref{sec:results} when comparing theory with simulations, this simple correction
qualitatively describes finite-size effects and captures in good approximation the location of phase transitions.

\section{Monte Carlo simulations in the unconstrained ensemble}
\label{sec:MC_unconstrained}

To perform the simulations, we consider a cubic box of side $L=V^{1/d}$ and scaled particle-coordinates $\mathbf{s}_i$ defined by
$\mathbf{q}_i=V^{1/d}\mathbf{s}_i$, $i=1,\dots,N$. In the unconstrained ensemble, the probability density of finding the system in a
particular $N$-particle configuration occupying a volume $V$ is given by~\cite{Latella_2021}
\begin{eqnarray}
\fl\mathcal{P}(N,V;\mathbf{s}^N)&=\frac{\beta P}{\Upsilon(\mu,P,T)} 
\exp\left[\beta\mu N-\beta PV+N\ln(V/\lambda_T^d)-\ln N!-\beta W(\mathbf{s}^N;V)\right],\nonumber\\
\label{prob_density}
\end{eqnarray}
where $\mathbf{s}^N\equiv(\mathbf{s}_1,\dots,\mathbf{s}_N)$ and which is the distribution to be sampled in the simulations. For a given
system configuration $\mathcal{C}$ and following the Metropolis scheme~\cite{Frenkel}, MC trial moves in this case consist of displacement
of particles, insertion and removal of particles and changes of volume, defining so a new configuration $\mathcal{C}'$. Below we denote
by $\Delta W=W(\mathcal{C}')-W(\mathcal{C})$ the variation of potential energy associated to the considered MC move.

In the case of displacement of particles, a trial move is attempted by selecting a particle at random and performing a random displacement
from $\mathbf{s}$ to $\mathbf{s}'$. From~(\ref{prob_density}) and according to the Metropolis rule, this move is accepted with a probability
\begin{equation}
P_\mathrm{acc}(\mathbf{s}\to \mathbf{s}')=
\min\left( 1, e^{-\beta\Delta W}\right).
\label{acceptance_particle_displacement}
\end{equation}
For the insertion of a particle, an attempt is made to add the particle at a random position keeping the remaining particles at the same
position and the volume fixed. Taking into account expression~(\ref{prob_density}), the acceptance probability in this case is
\begin{equation}
P_\mathrm{acc}(N\to N+1)=
\min\left[1, 
\frac{V e^{-\beta(\Delta W-\mu)}}{\lambda_T^d(N+1)}
\right].
\label{acceptance_particle_insertion}
\end{equation}
Similarly, for the removal of a particle chosen at random the acceptance probability takes the form
\begin{equation}
P_\mathrm{acc}(N\to N-1)=
\min\left[1, 
\frac{\lambda_T^d N}{V} e^{-\beta(\Delta W + \mu)}
\right].
\label{acceptance_particle_removal}
\end{equation}
Finally, from~(\ref{prob_density}), trial moves that attempt to perform a random increment of the volume from $V$ to $V'$ have an
acceptance probability given by
\begin{equation}
P_\mathrm{acc}(V\to V')=
\min\left[1, 
e^{N\ln(V'/V)
-\beta P(V'-V) -\beta \Delta W }
\right].
\label{acceptance_volume_displacement}
\end{equation}

The probabilities (\ref{acceptance_particle_displacement}), (\ref{acceptance_particle_insertion}) and (\ref{acceptance_particle_removal})
constitute the usual acceptance rules in the grand canonical ensemble, while (\ref{acceptance_particle_displacement}) and
(\ref{acceptance_volume_displacement}) are the acceptance probabilities in the isothermal-isobaric ensemble~\cite{Frenkel}. Thus, as
shown in~\cite{Latella_2021}, a consistent MC algorithm for simulations in the unconstrained ensemble can be obtained as a simple
combination of the algorithms for the grand canonical and isothermal-isobaric ensembles.

In the considered model, energies and temperatures are measured in units of the coupling constant $\nu$, which we set to $\nu=1$ in the
simulations. Moreover, $d$-dimensional volumes are measured in units of $V_0$, and we also set $V_0=1$. We furthermore particularize the
above acceptance probabilities in terms of the reduced volume $v$, pressure $p$ and chemical potential $\xi$ of the model defined in
equations (\ref{reduced_variables_1}) and (\ref{reduced_variables_2}). The acceptance probabilities (\ref{acceptance_particle_insertion})
and (\ref{acceptance_particle_removal}) for the insertion and removal of a particle become
\begin{equation}
P_\mathrm{acc}(N\to N+1)=
\min\left[1, \frac{T (v+1)}{(N+1)} e^{ -(\Delta W/T+\xi) } \right].
\label{acceptance_particle_insertion_reduced}
\end{equation}
and
\begin{equation}
P_\mathrm{acc}(N\to N-1)=\min\left[1,  \frac{ N }{T(v+1)} e^{ -(\Delta W/T-\xi) } \right],
\label{acceptance_particle_removal_reduced}
\end{equation}
respectively.
In the same way, using the reduced volume $v$, the acceptance probability (\ref{acceptance_volume_displacement}) takes the form
\begin{equation}
P_\mathrm{acc}(v\to v')= \min\left\{1, e^{
N\ln\left[ (v'+1)/(v+1)\right]  -Tp(v'-v) - \Delta W/T} \right\}.
\label{acceptance_volume_desplacement_reduced}
\end{equation}

To implement the algorithm, we generate a random integer $R$ such that $1\leq R\leq m$, where $m=N_\mathrm{av}+N_\mathrm{ex}+1$ is the
number of MC moves in a cycle, $N_\mathrm{av}$ and $N_\mathrm{ex}$ being fixed integers. We then attempt a particle displacement if
$R\leq N_\mathrm{av}$, a volume change if $R=N_\mathrm{av}+1$, and a particle exchange with the reservoir (insertion or removal with the
same probability) otherwise.  Accordingly, $N_\mathrm{av}$ particle displacements, $N_\mathrm{ex}$ particle exchanges and one volume change
are performed per cycle on average. In addition, here we take $N_\mathrm{ex}=N_\mathrm{av}$. Since the actual average number of
particles $\bar{N}$ is not known a priori, in a calibration stage we periodically set $N_\mathrm{av}=N$ so that $N_\mathrm{av}$ is
approximately $\bar{N}$. In this stage we also calibrate the maximum particle displacement and maximum volume variation to achieve an
acceptance ratio of about $50\%$. After calibration, we perform a thermalization run keeping all parameters fixed, compute the average
number of particles $\bar{N}$ and set $N_\mathrm{av}=\bar{N}$ when this run is finished. Lastly, in the production run we recompute
$\bar{N}$ as well as the other averaged quantities. In the simulations presented below, the total number of MC moves per particle is
$10^6$ in both the thermalization and production runs.

\section{Monte Carlo simulations in the isothermal-isobaric ensemble}
\label{sec:MC_isothermal}

While the total number of particles fluctuates in the unconstrained ensemble, here we focus on the situation in which $N=N_0+N_1$ is
fixed at constant pressure and temperature, as described by the isothermal-isobaric ensemble. In order to compare the simulations in the
unconstrained ensemble with those in the isothermal-isobaric case, we need to obtain the chemical potential in the latter. We do this by
following the Widom particle insertion method~\cite{Widom} in the $NPT$ ensemble~\cite{Shing,Sindzingre,Frenkel}.

The Gibbs free energy is given by
\begin{equation}
G(N,P,T)=-T\ln \left[\int \mathrm{d}V\ e^{-\beta PV}\int \frac{\mathrm{d}^{dN}\mathbf{q}}{\lambda_T^{dN}N!} \ e^{-\beta W(\mathbf{q}^N)}\right] .
\end{equation}
In this ensemble, the chemical potential can be obtained as $\mu=G(N+1,P,T)-G(N,P,T)$.
Since the free energy of a system with an additional particle can be written as
\begin{equation}
\fl G(N+1,P,T)=-T\ln\left[\int \mathrm{d}V\ e^{-\beta PV}
\int \frac{\mathrm{d}^{dN}\mathbf{q}}{\lambda_T^{dN}N!} \ e^{-\beta W(\mathbf{q}^{N})} 
\int \frac{\mathrm{d}^{d}\mathbf{q}_{N+1}}{\lambda_T^{d}(N+1)}
  \ e^{-\beta w}\right] ,
\end{equation}
the chemical potential can be computed as~\cite{Frenkel}
\begin{equation}
\mu=-T\ln\left\langle
\frac{1}{\lambda_T^{d}(N+1)} \int \mathrm{d}^{d}\mathbf{q}_{N+1} \ e^{-\beta w} \right\rangle,
\label{mu_widom}
\end{equation}
where we have introduced
$w\equiv W(\mathbf{q}^{N+1})-W(\mathbf{q}^{N})$.
Here $\langle \cdots \rangle$ indicates average in the isothermal-isobaric ensemble over the configuration space of the $N$-particle
system.  In our case, we can split the integral occuring in (\ref{mu_widom}) over the two regions with volumes $V_0$ and $V_1$, hence
\begin{equation}
\mu=-T\ln \left[\frac{V_0}{\lambda_T^{d}(N+1)}\right] -T\ln \left[ \left\langle I_0\right\rangle + \left\langle I_1\right\rangle \right],
\end{equation}
where
\begin{equation}
I_0=\frac{1}{V_0}\int_{V_0} \mathrm{d}^{d}\mathbf{q}_{N+1} \ e^{-\beta w},\qquad
I_1=\frac{1}{V_0}\int_{V_1} \mathrm{d}^{d}\mathbf{q}_{N+1} \ e^{-\beta w}. 
\end{equation}

To compute the averages $\langle I_0\rangle$ and $\langle I_1\rangle$, we exploit the fact that the regions of the system in the volumes
$V_0$ and $V_1$ are homogeneous (with different particle density, in general) and attempt the addition of a virtual particle at a random
position separately in each of the two regions.  In this way, we evaluate the integrals $I_0$ and $I_1$ individually, with the advantage
that the same $N$-particle configuration can be used for both cases. If the addition of the particle in region $k$ leads to an overlap
with the other particles, we have $I_k=0$ because the potential energy goes to infinity. If there is no overlap, we have
$W(\mathbf{q}^{N}) =-\nu [N_0(N_0-1) + b N_1(N_1-1)]$ and 
\begin{equation}
W(\mathbf{q}^{N+1})=\left\{
\begin{array}{cc}
-\nu [N_0(N_0+1) + b N_1(N_1-1)] & \mathrm{if}\ \mathbf{q}_{N+1} \in V_0 \\
-\nu [N_0(N_0-1) + b N_1(N_1+1)] & \mathrm{if}\ \mathbf{q}_{N+1} \in V_1 
\end{array}
\right. ,
\end{equation}
so that
\begin{eqnarray}
w&=\left\{
\begin{array}{cc}
-2\nu N_0 & \mathrm{if}\ \mathbf{q}_{N+1} \in V_0 \\
-2\nu b N_1 & \mathrm{if}\ \mathbf{q}_{N+1} \in V_1 
\end{array}
\right. \nonumber\\
\end{eqnarray}
and therefore
\begin{equation}
I_0=e^{2\beta\nu N_0},\qquad I_1=v e^{2b\beta \nu N_1},
\end{equation}
where we have used that $V_1/V_0=v$.
In this way, the reduced chemical potential in the isothermal-isobaric ensemble takes the form
\begin{equation}
\xi =\frac{\mu_T-\mu}{T}= \ln\left[ \left\langle I_0\right\rangle + \left\langle I_1\right\rangle \right]
-\ln\left[\nu (N+1)/T\right].
\end{equation}
This expression for the chemical potential allows for a direct comparison with the unconstrained ensemble in which $\xi$ is a control
parameter.

Finally, MC moves in this ensemble consist of particle displacement which are accepted with the probability
(\ref{acceptance_particle_displacement}) and volume variations whose acceptance probability is given by
(\ref{acceptance_volume_desplacement_reduced}). In this ensemble, we also take the total number of MC moves per particle equal
to $10^6$ in both the thermalization and production runs.

\section{Results}
\label{sec:results}

\begin{figure}
\centering
\includegraphics[scale=1]{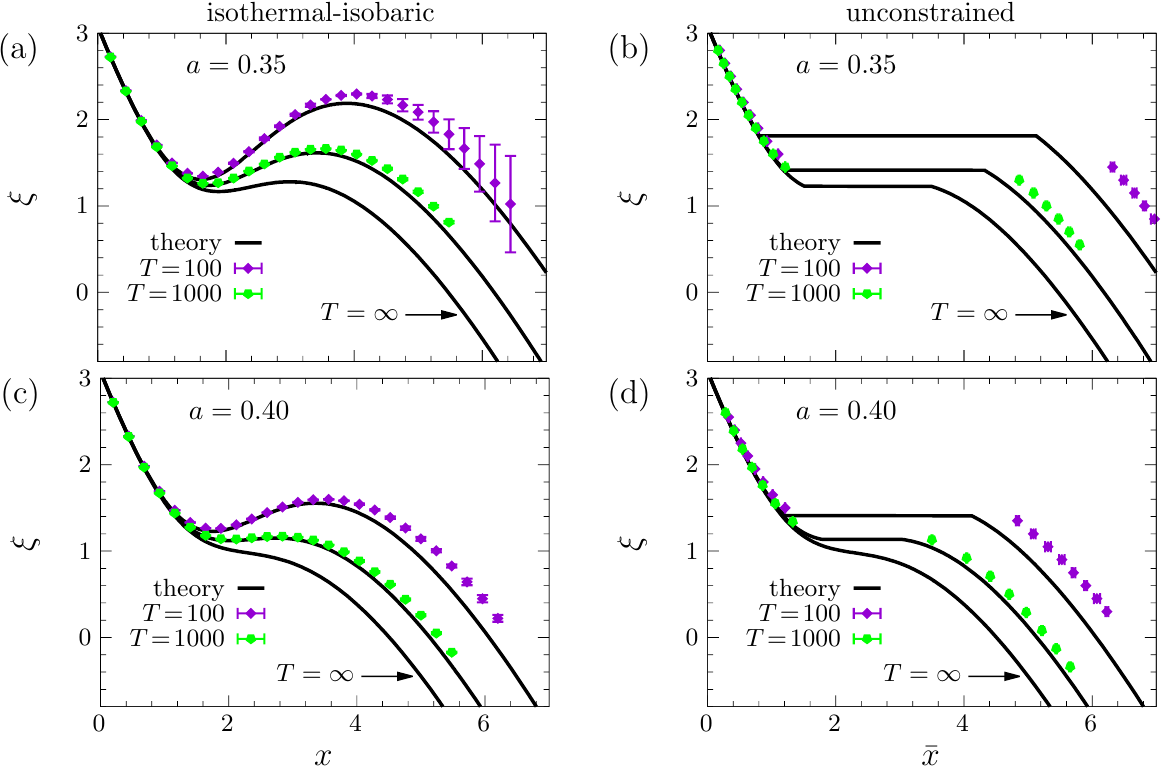}
\caption{MC simulations of the model in $d = 3$ for different temperatures. In (a) and (c) we show the reduced chemical potential as
a function of the reduced number of particles in the isothermal-isobaric ensemble, while in (b) and (d) we show the chemical potential
as a function of the average reduced number of particles in the unconstrained ensemble. In all cases the reduced pressure and coupling
strength are $p=0.05$ and $b=-1$, respectively. The exclusion parameter is $a=0.35$ in (a) and (b), whereas $a=0.40$ in (c) and (d).
The error bars show the standard deviation obtained from eight independent simulation runs.}
\label{xi_vs_x_conf_1_d3}
\end{figure}

\begin{figure}
\centering
\includegraphics[scale=1]{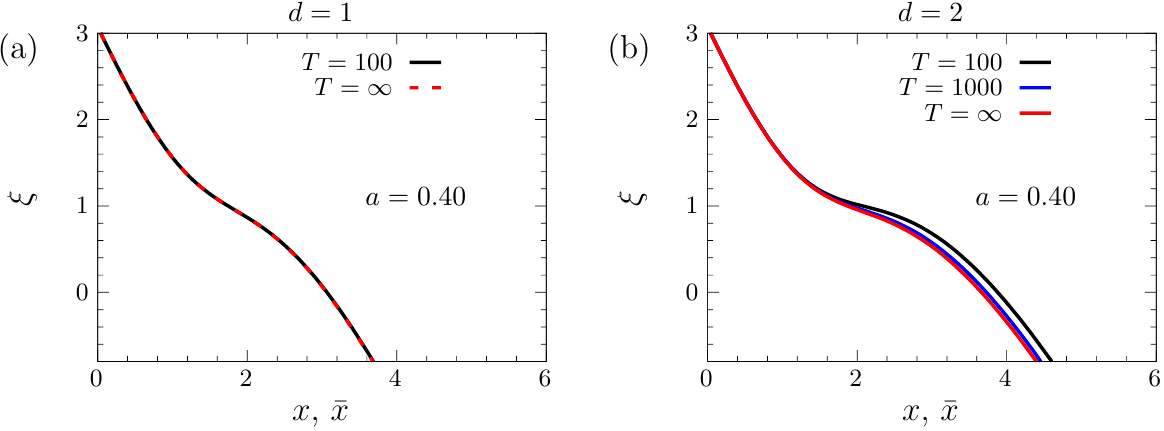}
\caption{Theoretical dependence on the finite-temperature correction for $d=1$ in (a) and $d=2$ in (b). 
The parameters are $p=0.05$, $a=0.40$ and $b=-1$. The unconstrained and isothermal-sobaric ensembles coincide in this regime.}
\label{theory_xi_vs_x_conf_1}
\end{figure}

In this section we present the results of  MC simulations performed in both the isothermal-isobaric and unconstrained ensembles for
different configurations of the model. The results in the plots represent an average over eight independent simulation runs and the
associated error bars show the corresponding standard deviations. A comparison with theoretical predictions is also made in which the
theoretical curves are all obtained with the finite-size correction (\ref{corrected_eta0}) for the packing fraction
in the core.

We first discuss the effect of taking finite temperature in the simulations.
Since the reduced total number of particles is given by $x=\nu N/ T$ (with the energy scale $\nu=1$ in the simulations), the limit
$T\to \infty$ in this case corresponds to the large $N$ limit with $x$ fixed. Therefore, finite temperature here means a finite number
of particles. The actual number of particles or its average in the simulations is directly obtained by multiplying $x$ or $\bar{x}$ by $T$. 
Furthermore, in the derivation of the finite-size correction in section~\ref{sec:finite_temp_correction}, we showed that the associated
finite-size effects become more important as the dimension $d$ increases. To put this fact in evidence, on the one hand, we performed
simulations in $d=3$ at different temperatures that are presented in figure~\ref{xi_vs_x_conf_1_d3} for both the isothermal-isobaric and
unconstrained ensembles. Here we take $a=0.35$ in the upper panels and $a=0.40$ in the lower panels, with $p=0.05$ and $b=-1$ in both
of them. In the plots, we compare with the theoretical predictions taking into account the correction (\ref{corrected_eta0}) for
finite-temperature effects, and we also include the plots for $T\to\infty$ corresponding to the case without this correction. The curves
correctly describe the diluted phase (at small $x$ or $\bar{x}$), qualitatively characterize the collapsed phase (at large $x$ or
$\bar{x}$) and remarkably account for the occurence of the phase transitions in the unconstrained ensemble which are identified by a sudden
increase in $\bar{x}$ at some chemical potential $\xi$. The counterpart in the isothermal-isobaric ensemble of these phase transitions is
the appearence of a portion of the curve $\xi(x)$ with positive slope (a possibility justified in section~\ref{theory_iso_ensemble}), as
can be seen in the simulations of figures~\ref{xi_vs_x_conf_1_d3}(a) and~\ref{xi_vs_x_conf_1_d3}(c). Notice that the phase transition in
figure~\ref{xi_vs_x_conf_1_d3}(b) for $a=0.35$ is still present in the case $T\to\infty$, while it is absent for $a=0.40$ in
figure~\ref{xi_vs_x_conf_1_d3}(d) for this temperature limit. Finite-temperature effects then shift the location of phase transitions in
the phase diagram. Moreover, as expected, the agreement between theory and simulations improves with increasing temperature. On the other
hand, the finite-temperature correction is relatively small in $d=2$ and almost negligible in $d=1$, as the theoretical predictions show
in figure~\ref{theory_xi_vs_x_conf_1} for an exclusion parameter $a=0.40$. In figure~\ref{xi_vs_x_conf_1} and in
figure~\ref{x0_x1_v_conf_1}, see below, we will compare the theoretical and the simulation results for all the different dimensions $d=1,2,3$
at $T=1000$.

The simulations clearly indicate that the considered ensembles are not equivalent for some range of control parameters.
As noted before, a region with positive slope in the curve $\xi(x)$ can be realized in the isothermal-isobaric ensemble. In the
unconstrained ensemble, the response function
\begin{equation}
M_{T,P}=\left(\frac{\partial \bar{N}}{\partial \mu}\right)_{T,P}=-\left(\frac{\partial^2 \mathscr{E}}{\partial \mu^2}\right)_{T,P}\geq0
\end{equation}
cannot be negative, as can be deduced from curvature properties of the replica energy~\cite{Campa_2018} (see also \cite{Hill_2013}). 
This is what one expects for a situation in which the number of particles is allowed to fluctuate. 
For the model considered here, $M_{T,P}$ can be written in terms of the reduced variables (\ref{reduced_variables_1}) and
(\ref{reduced_variables_2}), from which we infer that
\begin{equation}
\left(\frac{\partial \bar{x}}{\partial \xi}\right)_{T,p} \leq 0
\end{equation}
in the unconstrained ensemble. As a consequence, states associated with $\partial \xi/\partial x>0$ in the isothermal-isobaric ensemble
are jumped over by a first-order phase transition in the unconstrained ensemble.

\begin{figure}[!t]
\centering
\includegraphics[scale=1]{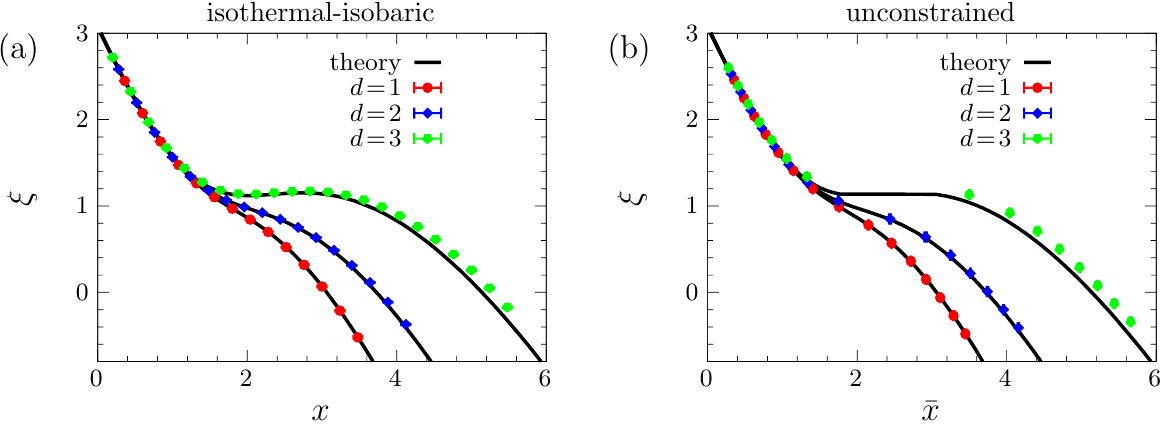}
\caption{MC simulations of the model for different space dimensions. In (a) we show the reduced chemical potential as a function of the
number of particles in the isothermal-isobaric ensemble, while in (b) we show the chemical potential as a function of the average reduced
number of particles in the unconstrained ensemble. In all cases the parameters are $p=0.05$, $a=0.40$, $b=-1$ and $T=1000$.}
\label{xi_vs_x_conf_1}
\end{figure}
\begin{figure}[!t]
\centering
\includegraphics[scale=1]{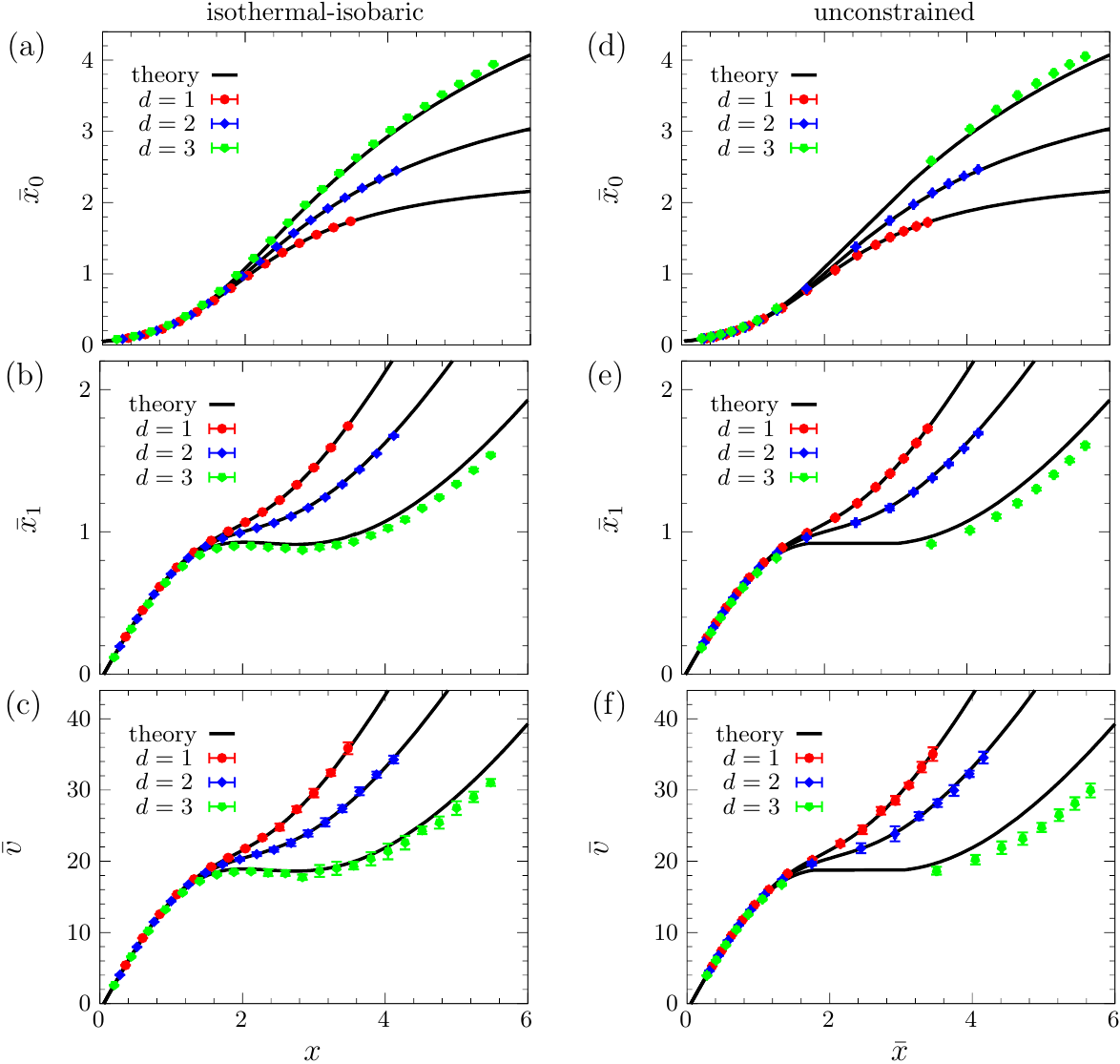}
\caption{Reduced volume and reduced number of particles in the two internal regions as a function of the total reduced number of
particles, which corresponds to the same simulations shown in figure~\ref{xi_vs_x_conf_1}. In all cases the parameters are $p=0.05$,
$a=0.40$, $b=-1$ and $T=1000$.}
\label{x0_x1_v_conf_1}
\end{figure}

In figure~\ref{xi_vs_x_conf_1} we show the relation between reduced chemical potential and number of particles in the isothermal-isobaric
and unconstrained ensembles in $d=1,2,3$ spatial dimensions and with parameters $p=0.05$, $a=0.40$, $b=-1$ and $T=1000$. In these simulations,
the average number of particles ranges, for instance, from $\bar{N}\approx 280$ to $\bar{N}\approx 5600$ in the unconstrained ensemble
for $d=3$, and a similar range for $N$ is taken in the $NPT$ ensemble. The simulation data, as in the previous figures, are compared
with the theory; the agreement is very satisfactory, implying in particular that the theoretical treatment takes care with good accuracy
of the finite-size corrections. For these parameter values the data show that a phase transition is present for $d=3$, but not for $d=1$ and
$d=2$, confirming that the increase of the dimension facilitates the occurrence of a phase transition.
For the same simulations shown in figure~\ref{xi_vs_x_conf_1}, in figure~\ref{x0_x1_v_conf_1}, we represent $\bar{x}_0$, $\bar{x}_1$ and
$\bar{v}$, again comparing with the thoretical evaluations. We highlight that the jump in $\bar{x}$ in the unconstrained
ensemble is due to a jump in the number of particles in the core $\bar{x}_0$, while the number of particles and volume of the outer
region $\bar{x}_1$ and $\bar{v}$, respectively, approach the same value in the two phases. This can be seen by comparing the $d=3$
data in the three plots in the right column of figure~\ref{x0_x1_v_conf_1}.

\begin{figure}
\centering
\includegraphics[scale=1]{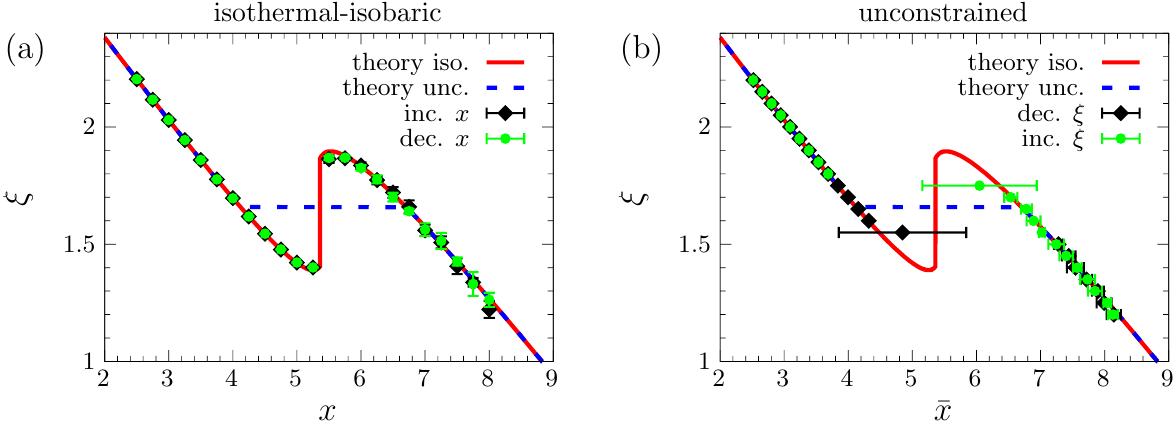}
\caption{MC simulations of the model in $d=1$. In (a) we show the reduced chemical potential as a function of the number of particles in
the isothermal-isobaric ensemble, while in (b) we show the chemical potential as a function of the average reduced number of particles in
the unconstrained ensemble. In both cases the parameters are $p=0.045$, $a=0.23$, $b=-3/16$ and $T=100$. Large error bars in (b) at the phase
transition reflects the fact that different realizations were obtained in either the diluted phase or the collapsed phase.}
\label{conf_paper}
\end{figure}

In the previous simulations, we have chosen the control parameters such that a phase transition is realized only in the unconstraned ensemble.
As shown in~\cite{Campa_2020}, the model also exhibits first-order phase transitions in the isothermal-isobaric ensemble.
In figure~\ref{conf_paper} we show the simulation of a configurations shown in~\cite{Campa_2020} derived from the excluded-volume
approximation which, as we discussed previously, is exact in $d=1$. For these simulations, we set $p=0.045$, $a=0.23$ and $b=-3/16$ in the
two ensembles. In this figure we show the results obtained by increasing and decreasing the number of particles in the isothermal-isobaric
ensemble and by decreasing and decreasing the chemical potential in the unconstrained ensemble. The phase transition in the unconstrained
ensemble exhibit hysteresis, as clearly observed in figure~\ref{conf_paper}(b). Moreover, different realizations of the simulations around
this transition were found in either the diluted or the collapsed phase, which in the plot is represented by relatively large error bars. 

\section{Discussion}
\label{sec:discussion}

We have studied equilibrium states of a modified version of the Thirring model with attractive and repulsive long-range interactions in
which particles are treated as hard spheres. By developing a theoretical framework and performing MC simulations, we have shown that the
model presents first-order phase transitions under completely open conditions, in the unconstrained ensemble. In our simulations we have
also verified that the unconstrained and the isothermal-isobaric ensembles are not equivalent in this model.

The model considered here was previously analyzed in~\cite{Campa_2020} from a theoretical point of view, treating the hard-sphere
interactions in the excluded-volume approximation. While this approach is exact in the thermodynamic limit for $d=1$ spatial dimensions,
it fails to accurately describe relatively high density states and the location of phase transitions for $d>1$. To go beyond this
approximation, we developed a theoretical description for an arbitrary equation of state and dimensionality that becomes exact in the
thermodynamic limit. By studying some system configurations for different spatial dimensions, we confronted this theoretical description
with MC simulations and considered finite-size effects with an approximate correction. This finite-size correction adequately captures the
location of phase transitions in the simulations. 

Our work highlights the rich phenomenology displayed by long-range interacting systems. In particular, the possibility of observing
equilibrium states under completely open conditions in which energy, volume and number of particles fluctuate.

Completely open conditions in this work, in particular the exchange of heat with the surroundings of the system, have been treated
with a MC scheme. In another approach the exchange of heat between self-gravitating systems and the surroundings was modeled
with a Smoluchowski equation~\cite{Chavanis_2002_c}, showing the different behavior with respect to the isolated system.
Concerning ensemble inequivalence, it was shown
that inequivalence between the microcanonical and the canonical ensembles occurs in self-gravitating fermions~\cite{Chavanis_2006},
in which the Pauli exclusion principle physically plays the role of the short-range repulsion, due in our case to the finite size
of the particles. A phenomenolgy similar to ours was found with the inverse temperature $\beta$ and the energy
playing analogous roles of our parameter $\xi$, the reduced chemical potential, and our parameter $x$, the reduced
total number of particles, respectively. Here we have shown that the introduction of a long-range repulsion ($b<0$)
together with a finite particle size can lead not only to phase transitions in the unconstrained ensemble~\cite{Campa_2020},
but also to inequivalence of this enemble with that in which the number of particles is constrained.

For non-additive systems in which both the long-range part and the short-range part of the interaction are more general than
those of the model studied in this paper, in particular when the short-range component is represented by a smooth potential, it
will in general more difficult, or even unachievable, to find a theoretical representation of the finite-size effects giving
results that are quantitatively satisfactory. The search of the size-dependent location of the phase transitions has in this case to rely
on the simulations, and in this respect it is very useful to have a tool like the MC scheme employed in this work. It can be used in
all circumstances in which a system can reach equilibrium states under completely open conditions, and we hope that it will find
various applications.

\section*{Acknowledgments}
A. C. acknowledges financial support from INFN (Istituto Nazionale di Fisica Nucleare) through the projects DYNSYSMATH and ENESMA.
J. M. R. acknowledges financial support from MICIU (Spanish Government) Grant No. PGC2018-098373-B-I00. This work is part of the
MIUR-PRIN2017 project Coarse-grained description for nonequilibrium systems and transport phenomena (CO-NEST) No. 201798CZL.


\section*{References}
\begin{thebibliography}{50}

\bibitem{Campa_2014} Campa A, Dauxois T, Fanelli D and Ruffo S
2014 {\it Physics of Long-Range Interacting Systems} (Oxford: Oxford University Press)

\bibitem{Campa_2009} Campa A, Dauxois T and Ruffo S
009 {\it Phys. Rep.} {\bf 480} 57

\bibitem{Levin_2014} Levin Y, Pakter R, Rizzato F B, Teles T N and Benetti F P C
2014 {\it Phys. Rep.} {\bf 535} 1

\bibitem{Bouchet_2010} Bouchet F, Gupta S and Mukamel D
2010 {\it Physica} A {\bf 389} 4389

\bibitem{Feliachi_2022} Feliachi O and Bouchet F 
2022 {\it J. Stat. Phys.} {\bf 186} 22

\bibitem{Nicholson_1992} Nicholson D R
1992 {\it Introduction to Plasma Physics} (Malabar, FL: Krieger)

\bibitem{Kiessling_2003} Kiessling M K H and Neukirch T
2003 {\it Proc. Natl. Acad. Sci.} {\bf 100} 1510

\bibitem{Onsager_1949} Onsager L
1949 {\it Nuovo Cimento Suppl.} {\bf 6} 279

\bibitem{Miller_1990} Miller J
1990 {\it Phys. Rev. Lett.} {\bf 65} 2137

\bibitem{Robert_1991} Robert R and Sommeria J
1991 {\it J. Fluid. Mech.} {\bf 229} 291

\bibitem{Chavanis_2002_b} Chavanis P-H and Sommeria J
2002 {\it Phys. Rev.} E {\bf 65} 026302

\bibitem{Bouchet_2009} Bouchet F and Simonnet E
2009 {\it Phys. Rev. Lett.} {\bf 102} 094504

\bibitem{Venaille_2012} Bouchet F and Venaille A
2012 {\it Phys. Rep.} {\bf 515} 227

\bibitem{Barre_2004} Barr\'e J, Dauxois T, De Ninno G, Fanelli D and Ruffo S 
2004 {\it Phys. Rev. E} {\bf 69} 045501(R)

\bibitem{Barre_2005} Barr\'e J, Bouchet F, Dauxois T and Ruffo S
2005 {\it J. Stat. Phys.} {\bf 119} 677

\bibitem{Antonov_1962} Antonov V A
1962 Vest. Leningr. Gos. Univ. {\bf 7} 135 
\\ Antonov V A 1985 {\it IAU Symposia} {\bf 113} 525 (translation)

\bibitem{Lynden-Bell_1968} Lynden-Bell D and Wood R
1968 {\it Mon. Not. R. Astr. Soc.} {\bf 138} 495

\bibitem{Thirring_1970} Thirring W
1970 {\it Z. Phys.} {\bf 235} 339

\bibitem{Padmanabhan_1990} Padmanabhan T
1990 {\it Phys. Rep.} {\bf 188} 285

\bibitem{Lynden-Bell_1999} Lynden-Bell D
1999 {\it Physica A} {\bf 263} 293

\bibitem{Chavanis_2002} Chavanis P-H
2002 {\it Astron. Astrophys.} {\bf 381} 340

\bibitem{Chavanis_2006} Chavanis P-H
2006 {\it Int. J. Mod. Phys.} B {\bf 20} 3113

\bibitem{Ellis_2000} Ellis R S, Haven K and Turkington B
2000 {\it J. Stat. Phys.} {\bf 101} 999

\bibitem{Barre_2001} Barr\'e J, Mukamel D and Ruffo S
2001 {\it Phys. Rev. Lett.} {\bf 87} 030601

\bibitem{Bouchet_2005} Bouchet F and Barr\'e J
2005 {\it J. Stat. Phys.} {\bf 118} 1073

\bibitem{Latella_2015} Latella I, P\'erez-Madrid A, Campa A, Casetti L and Ruffo S
2015 {\it Phys. Rev. Lett.} {\bf 114} 230601

\bibitem{Latella_2013} Latella I and P{\'e}rez-Madrid A
2013 {\it Phys. Rev.} E {\bf 88} 042135

\bibitem{Latella_2017} Latella I, P\'erez-Madrid A, Campa A, Casetti L and Ruffo S
2017 {\it Phys. Rev.} E {\bf 95} 012140

\bibitem{Hill_2013} Hill T L
2013 {\it Thermodynamics of Small systems, Parts I and II} (New York: Dover)

\bibitem{Hill_2001} Hill T L
2001 {\it Nano Lett.} {\bf 1} 273

\bibitem{Frenkel} Frenkel D and Smit B
2002 \textit{Understanding Molecular Simulation: From Algorithms to Applications} (San Diego: Academic Press)

\bibitem{Latella_2021} Latella I, Campa A, Casetti L, Di Cintio P, Rubi J M and Ruffo S
2021 {\it Phys. Rev.} E {\bf 103} L061303

\bibitem{Campa_2016} Campa A, Casetti L, Latella I, P\'erez-Madrid A and Ruffo S
2016 {\it J. Stat. Mech.} 073205

\bibitem{Trugilho_2022} Trugilho L F and Rizzi L G
2022 {\it J. Stat. Phys.} {\bf 186} 40

\bibitem{Campa_2020} Campa A, Casetti L, Latella I and Ruffo S
2020 {\it J. Stat. Mech.} 014004

\bibitem{Aronson_1972} Aronson E B and Hansen C J
1972 {\it Astrophys. J.} {\bf 177} 145

\bibitem{Tonks_1936}Tonks L
1936 {\it Phys. Rev.} {\bf 50} 955 

\bibitem{Kac_1963} Kac M, Uhlenbeck G E and Hemmer P
1963 {\it J. Math. Phys.} {\bf 4} 216-228

\bibitem{Chavanis_2011} Chavanis P-H
2011 {\it Physica} A {\bf 390} 1546

\bibitem{Chavanis_2019} Chavanis P-H
2019 {\it Entropy} {\bf 21} 1006

\bibitem{Baldovin_1} Baldovin F, Orlandini E 
2006 {\it Phys. Rev. Lett.} {\bf 96} 240602

\bibitem{Baldovin_2} Baldovin F, Orlandini E 
2006 {\it Phys. Rev. Lett.} {\bf 97} 100601 

\bibitem{Chavanis_2006_b} Chavanis P-H
2006 {\it Physica} A {\bf 361} 55

\bibitem{Hill_Statistical_Mechanics}Hill T L 
1956 \textit{Statistical Mechanics: Principles and Selected Applications} (New York: McGraw-Hill)

\bibitem{Chavanis_2020} Chavanis P-H
2020 {\it Eur. Phys. J. Plus} {\bf 135} 290

\bibitem{Henderson_1975} Henderson H
1975 {\it Mol. Phys.} {\bf 30} 971

\bibitem{Carnahan_1969} Carnahan N F and Starling K E 
1969 {\it J. Chem. Phys.} {\bf 51} 635

\bibitem{Santos_1995} Santos A, L\'opez de Haro M and Bravo Yuste S
1995 {\it J. Chem. Phys.} {\bf 103} 4622

\bibitem{Robles_2014} Robles M, L\'opez de Haro M and Santos A
2014 {\it J. Chem. Phys.} {\bf 140} 136101

\bibitem{Widom} Widom B 
1963 {\it J. Chem. Phys.} {\bf 39} 2808

\bibitem{Shing} Shing K S
1985 {\it Chem. Phys. Lett} {\bf 119} 149

\bibitem{Sindzingre} Sindzingre P, Ciccotti G, Massobrio C and Frenkel D
1987 {\it Chem. Phys. Lett.} {\bf 136} 35

\bibitem{Campa_2018} Campa A, Casetti L, Latella I, P\'erez-Madrid A and Ruffo S
2018 {\it Entropy} {\bf 20} 907

\bibitem{Chavanis_2002_c} Chavanis P-H
2002 {\it Phys. Rev.} E {\bf 66} 036105

\end{thebibliography}
\end{document}